# EFFICIENCY OF HIGHER ORDER FINITE ELEMENTS FOR THE ANALYSIS OF SEISMIC WAVE PROPAGATION

J.F Semblat, J.J.Brioist

*Laboratoire Central des Ponts et Chaussées, Eng. Modelling Dept, 58, bd Lefèbvre, 75732 Paris Cedex 15, France (semblat@lcpc.fr)*

## 1. INTRODUCTION

The analysis of wave propagation problems in linear damped media must take into account both propagation features and attenuation process [1,3,6,10]. To perform accurate numerical investigations by the finite differences or finite element method, one must consider a specific problem known as the *numerical dispersion of waves*. Numerical dispersion may increase the numerical error during the propagation process as the wave velocity (phase and group) depends on the features of the numerical model [2,12]. In this paper, the numerical modelling of wave propagation by the finite element method is thus analyzed and discussed for linear constitutive laws. Numerical dispersion is analyzed herein through 1D computations investigating the accuracy of higher order 15-node finite elements towards numerical dispersion. Concerning the numerical analysis of wave attenuation, a rheological interpretation of the classical Rayleigh assumption has for instance been previously proposed in this journal [10].

## 2. WAVE PROPAGATION AND DISPERSION

### 2.1 NUMERICAL MODELLING OF SEISMIC WAVE PROPAGATION

Different types of numerical methods are available to investigate seismic wave propagation : finite differences, finite elements, spectral methods or boundary elements [3,5,6,13]. The main advantage of the boundary element method is to allow an accurate modelling of wave propagation in (semi-)infinite media. The finite element method is very efficient for the response analysis of complex non linear media. For the analysis of seismic wave propagation, the two main drawbacks of the finite element method are the artificial reflections on the mesh boundaries and the numerical dispersion. This paper considers the modelling of wave propagation problems through the second issue. Numerical dispersion is analyzed for 1D models and different types of finite elements (from low to higher order).

## 2.2 THEORETICAL AND PHYSICAL POINTS OF VIEW

For a viscoelastic solid, the one-dimensional wave equation in the frequency domain can be written as follows :

$$\frac{\partial^2 u(x,\omega)}{\partial x^2} + \frac{\rho \omega^2}{E^*(\omega)} u(x,\omega) = 0 \tag{1}$$

where $u$ is the displacement, $x$ the distance, $\omega$ the circular frequency, $\rho$ the density and $E^*(\omega)$ the complex modulus [1,11].

The solution can then take the following form [1,3] :

$$u(x,\omega) = u(0,\omega).\exp\left(ik^*(\omega)x\right) \tag{2}$$

where $k^*(\omega)$ is the complex wavenumber such as :

$$k^*(\omega) = \frac{\omega}{c(\omega)} + i\alpha(\omega) \tag{3}$$

The first term is related to the phase difference and the wave velocity $c(\omega)$ depends on frequency. This dependence characterizes the physical dispersion. The second term of equation 3 corresponds to damping and gives a real valued decreasing exponential term in the expression of solution (2). From the numerical point of view, both properties have their counterparts generally called *numerical dispersion* and *numerical damping* [6,12]. Numerical dispersion make the wave velocity depends on the features of the numerical model (time integration scheme, mesh size, element type...).

## 3. NUMERICAL WAVE DISPERSION

The physical (and geometrical) wave dispersion makes the wave velocity depend on frequency. *Numerical dispersion* makes the wave velocities change with the features of the numerical model. Propagation phenomena could then be difficult to model using finite difference or finite element methods since the numerical error may increase during propagation.

The numerical solution of equation 1 can be written under the same form than theoretical solution 2 :

$$u_h(x,\omega) = u(0,\omega).\exp\left(ik_h(\omega)x\right) \tag{4}$$

where $u_h$ and $k_h$ are the approximated displacement and wavenumber.

Different theoretical works are dealing with the estimation of the numerical error made on $k_h$ when compared with the exact wavenumber $k$ [2,8]. Ihlenburg and Babuška [8] give for instance the following relation for finite elements with linear interpolation :

$$\cos k^h h = \frac{1 - K^2/3}{1 + K^2/6} \qquad (5)$$

where $K$ is the normalized frequency such as $K=kh=\omega h/c$.

Expression 5 shows that the numerical solution of equation 1 is only a propagating wave for normalized frequencies below the cutoff frequency $K_0$ [8]. For such frequencies, the numerical wave is nevertheless propagating slower or faster than the theoretical solution, depending on the characteristics of the numerical model. One must analyze this numerical dispersion and quantify the corresponding error.

## 4. EFFICIENCY OF HIGHER ORDER FINITE ELEMENTS

### 4.1 NUMERICAL DISPERSION FOR LOW ORDER FINITE ELEMENTS

To analyze the numerical error for wave propagation problems, we have previsouly considered a simple one-dimensional case involving a linear elastic medium (no physical dispersion) and low order finite elements (linear polynomial interpolation) [12]. The numerical wave dispersion is investigated considering the ratio $\Delta h/\lambda$ which is the normalized size of the elements towards the wavelength $\lambda$. From these results, it can be noticed that coarse meshes lead to numerical results overestimating velocities (phase or group). This is the pratical effect of numerical dispersion which can be overwhelmed by using an element size well-adapted to the wavelength of the problem. Classically, the element size is chosen around a tenth or a twentieth of the wavelength. However, even with these assumptions, the numerical error may be significant for large propagating distances (for instance $5\lambda$ or $10\lambda$).

In two dimensions, it is necessary to take into account the wave type, the angle of incidence, the type of element (triangular, quadrilateral...). Bamberger et al. [2,12] give different dispersion relations for numerical waves through phase and group velocities. From these dispersion laws, several general conclusions for meshes with linear finite elements can be made :
− numerical dispersion is higher for a larger element size (compared to the wavelength) ;
− the error is maximum for a zero incidence and minimum for a 45° incidence angle ;
− for small element size values, P-waves are much more sensitive to incidence angle than S-waves.

For an element size to wavelength ratio of 0.5($\Delta h=\lambda/2$), the relative error on phase velocity can reach 50% for a quadrilateral elements mesh and 30% for a mesh involving triangular elements [2,8,12]. Whereas, for a value of 0.1 ($\Delta h=\lambda/10$), the relative error on phase velocity is below 2%. Using elements corresponding to a tenth or a twentieth of the wavelength leads to results of good precision. These are the usual values taken for pratical computations.

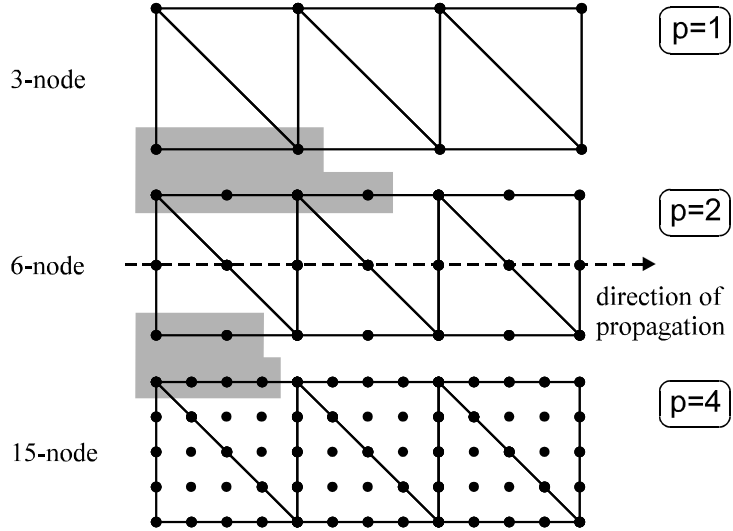

Figure 1. Different types of finite elements considered in the analysis with ratio of the number of nodes in the direction of propagation and degree of polynomial interpolation *p*.

TABLE 1

*Comparisons for different finite element orders and various number of elements.*

| Element type | 3-node | 6-node | 15-node |
| --- | --- | --- | --- |
| Case 1 | 200 | 100 | 50 |
| Case 2 | 120 | 60 | 30 |
| Case 3 | 80 | 40 | 20 |

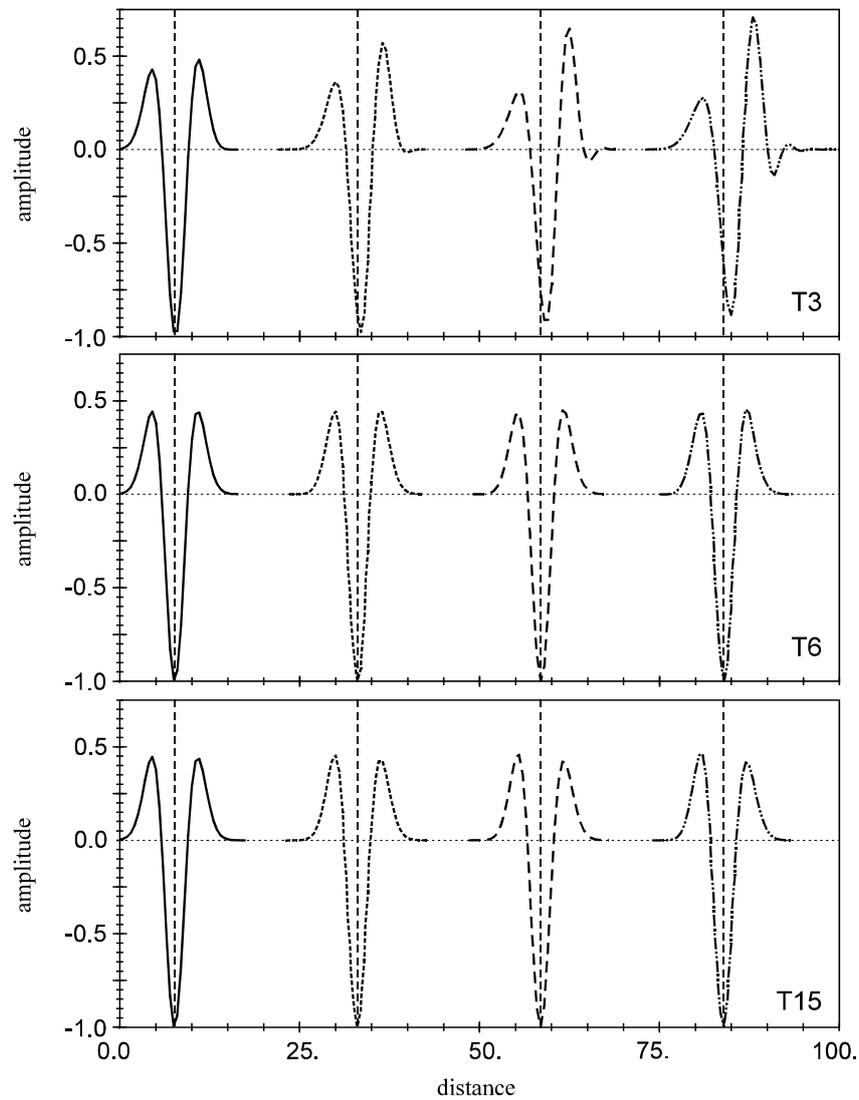

Figure 2 : Numerical dispersion (case 1) considering different types of finite elements : snapshots at different times and theoretical delays (computed with CESAR-LCPC).

4.2 COMPARISON OF DIFFERENT FINITE ELEMENT TYPES

The great interest of higher order finite elements has already been demonstrated for stress analysis involving elasto-plastic materials [9,14]. In this section, we analyze the efficiency of different element types towards numerical wave dispersion. Different types of finite elements are depicted in figure 1 from linear (3-node), quadratic (6-node) and higher order 15-node elements [6]. To make valuable comparisons, we study the one-dimensional wave propagation problem considering the same number of nodes in the direction of propagation for each type of element. A Newmark time integration algorithm (inconditionnally stable) is considered within the finite element code CESAR-LCPC [7].

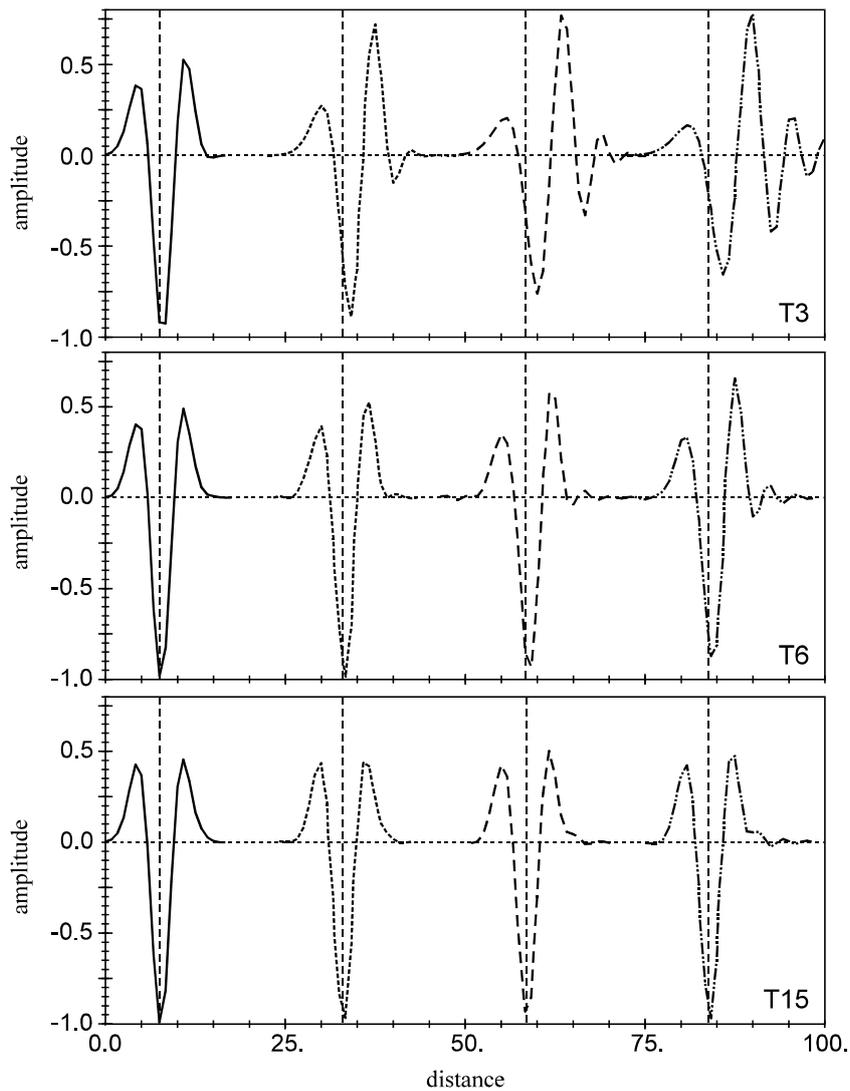

Figure 3 : Numerical dispersion (case 2) considering different types of finite elements : snapshots at different times and theoretical delays (computed with CESAR-LCPC).

Three different cases are studied ranging from rather fine to very coarse meshes. The total number of points in the direction of propagation is chosen constant from one element type to another (figure 1, table 1). The number of elements is then two times smaller for 6-node elements than for 3-node and four times smaller for 15-node elements. As the interpolation degrees for each element type are respectively 1, 2 and 4, the number of elements in each case is inversely proportional to the order of the polynomial approximation. The efficiency of these finite elements towards numerical wave dispersion is thus analyzed in terms of the ratio $\Delta h/n\lambda$ where $n$ is the degree of their polynomial interpolation.

4.3 EFFICIENCY TOWARDS NUMERICAL DISPERSION

Figures 2, 3 and 4 gives the numerical results for a second order Ricker pulse propagating in a linear elastic medium (no physical dispersion). These figures respectively correspond to cases 1, 2 and 3 of table 1 (from moderate to strong numerical dispersion). For linear elements (3-node), the numerical dispersion is already significant in case 1, is rather strong in case 2 and is very strong in case 3. For quadratic 6-node elements, there is no dispersion in case 1 and they appear more efficient than 3-node elements. In case 2 and case 3, 6-node elements nevertheless lead to significant and rather strong (resp.) numerical wave dispersion. In both first cases (1 and 2), the efficiency of higher order 15-node elements is very good since there is no numerical dispersion. For figure 4, numerical dispersion is very strong for linear elements, significant for quadratic elements and rather small for higher order elements. For case 3, some spurious oscillations nevertheless appear showing that the corresponding meshes are not fine enough (towards the wavelength involved).

Considering the same number of degrees of freedom in the direction of propagation, the accuracy and efficiency of higher order finite elements appear much better than linear 3-node and even quadratic 6-node elements. Ihlenburg and Babuška [8] also give some analytical estimation of the numerical error on wave velocity for different type of finite elements.

5. CONCLUSION

For wave propagation problems, the estimation of wave velocity is affected by some error called *numerical dispersion* and depending on many parameters such as mesh refinement, time integration scheme, element type... The classical rule is to choose the element size between a tenth and a twentieth of the wave-length. As the numerical error increases during propagation, it could not be sufficient to analyze far field wave propagation.

Higher order finite element are found to have a much better efficiency towards numerical dispersion than linear and even quadratic elements. However, the mesh refinement has to be sufficient to avoid spurious oscillations in the propagated wave. It is also necessary to consider the dispersive features of the time-integration scheme. For the analysis of seismic wave propagation, one must also investigate damping through both numerical and physical damping [1,3,10,11].

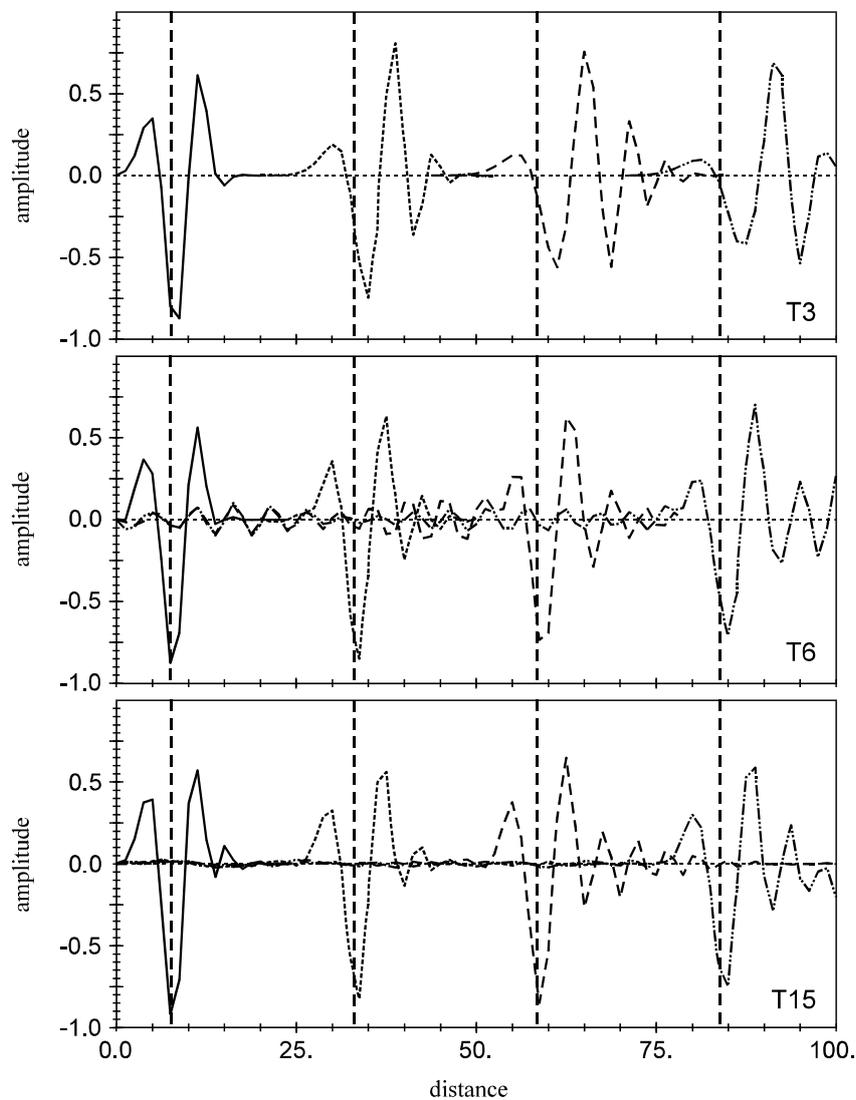

Figure 4 : Numerical dispersion (case 3) considering different types of finite elements : snapshots at different times and theoretical delays (computed with CESAR-LCPC).

## REFERENCES

1. K. Aki and P.G. Richards 1980 *Quantitative seismology.* San Francisco: Freeman ed.
2. A. Bamberger, G. Chavent and P. Lailly 1980. *Analysis of numerical schemes for linear elastodynamics equations* (in french), Report No.41, INRIA.
3. M. Bonnet 1999. *Boundary integral equation methods for solids and fluids.* Chichester, United Kingdom: Wiley.
4. T. Bourbié, O. Coussy and B. Zinszner 1986. *Acoustics of porous media.* Paris, France: Technip-Institut Français du Pétrole.
5. E. Faccioli, F. Maggio, R. Paolucci and A. Quarteroni 1997. *Journal of Seismology.* **1**, 237-251. 2D and 3D elastic wave propagation by a pseudo-spectral domain decomposition method.
6. T.J.R. Hughes 1987. *Linear static and dynamic finite element analysis.* Englewood Cliffs, New Jersey: Prentice-Hall.
7. P. Humbert 1989. *Bulletin des Laboratoires des Ponts et Chaussées* **160**, 112-115. CESAR-LCPC : a general finite element code (in French).
8. F. Ihlenburg and I. Babuška 1995. *Int. J. for Numerical Methods in Eng.* **38**, 3745-3774. Dispersion analysis and error estimation of Galerkin finite element methods for the Helmholtz equation.
9. F. Molenkamp and S. Kay 1997. *Numerical Models in Geomechanics*, 415-419. Comparison of convergence of 8-node and 15-node elements.
10. J.F. Semblat 1997. *Journal of Sound and Vibration* **206**(5), 741-744. Rheological interpretation of Rayleigh damping.
11. J.F. Semblat and M.P. Luong 1998. *Journal of Earthquake Engineering* **2**(1), 147-172. Wave propagation through soils in centrifuge testing.
12. J.F. Semblat 1998. *Revue Française de Génie Civil* **2**(1), 91-111. Waves attenuation and dispersion : physical and numerical points of view (in french).
13. J.F. Semblat, P. Dangla and A.M. Duval 1999. *2nd Int. Conf. on Earthquake Geotechnical Engineering*, 211-216. Seismic waves amplification : experimental versus numerical analysis, Balkema ed.

14. S.W. Sloan and F.M. Randolph 1982. *Int. J. for Num. and Analytical Meth. in Geomechanics*, **6**, 47-76. Numerical prediction of collapse loads using finite element methods.